\documentclass[aps,pra,twocolumn,groupedaddress,superscriptaddress,tightenlines,nofootinbib]{revtex4}
\usepackage{amssymb}
\usepackage{color,graphicx}
\usepackage{lmodern}
\usepackage{amsmath}
\usepackage{amsbsy}
\usepackage{float}
\usepackage{bbm}
\usepackage{bm}
\usepackage{epsfig}
\usepackage{braket}
\usepackage{graphics}
\usepackage{subfigure}
\usepackage{appendix}
\usepackage{lipsum}
\usepackage[T1]{fontenc}

\newcommand\D{\operatorname{d}}
\newcommand\EXP{\operatorname{exp}}

\newcommand\thh{\operatorname{th}}
\newcommand\Mod{\operatorname{mod}}
\newcommand\X{\operatorname{X}}
\newcommand\Y{\operatorname{Y}}

\newcommand\Rr{\bf{R}}
\newcommand\rR{\bf{r}}
\newcommand\Xx{\bf{x}}
\newcommand\sS{\bf{s}}

\begin{document}
\title{Molecular Electronic Structure Calculation via a Quantum Computer}

\author{Hamid Reza Naeij}
\email[]{naeij@alum.sharif.edu}
\affiliation{Iranian Quantum Technologies Research Center (IQTEC), Tehran, Iran}

\author{Erfan Mahmoudi}
\email[]{mahmoudi _ erfan@ch.sharif.edu}
\affiliation{Iranian Quantum Technologies Research Center (IQTEC), Tehran, Iran}

\author{Hossein Davoodi Yeganeh}
\email[]{h.yeganeh@ut.ac.ir}
\affiliation{Iranian Quantum Technologies Research Center (IQTEC), Tehran, Iran}

\author{Mohsen Akbari}
\email[Corresponding Author:~]{mohsen.akbari@khu.ac.ir}
\affiliation{Quantum Optics Lab, Department of Physics, Kharazmi University, Tehran, Iran}

\begin{abstract}

Quantum computers can be used to calculate the electronic structure and estimate the ground state energy of many-electron molecular systems. In the present study, we implement the Variational Quantum Eigensolver (VQE) algorithm, as a hybrid quantum-classical algorithm to calculate the ground state energy of the molecules such as H$^+_3$,  OH$^-$, HF and BH$_3$ in which the number of qubits has an increasing trend.  We use the parity transformation for fermion to qubit encoding and the Unitary Coupled Cluster for Single and Double excitations (UCCSD) to construct an ansatz. We compare our quantum simulation results with the computational chemistry approaches including Full Configuration Interaction (FCI), as benchmark energy and Unrestricted Hartree-Fock (UHF), as a common computational method. Our results show that there is a good agreement between molecular ground state energy obtained from VQE and FCI. Moreover, the accuracy of the ground state energies obtained from VQE in our work is higher than the previously reported values. This work aims to benchmark the VQE algorithm to calculate the electronic ground state energy for a new set of molecules that can be good candidates for molecular simulation on a real quantum computer.

\end{abstract}

\maketitle
\textbf{Keywords}: quantum computer, variational quantum algorithm, ansatz, ground state energy

\section{Introduction}

Quantum computer is a rapidly-emerging technology based on quantum features such as superposition and entanglement to solve problems that may not be suitable for classical computers \cite{Feynman,Mermin,Nielsen}. Quantum computer uses quantum algorithms to perform efficient calculations, resulting in an exponential speedup or improved efficiency compared to the classical algorithm. Shor's algorithm for integer factorization \cite{Shor} and Grover's algorithm for unstructured database search \cite{Grover} are two famous algorithms in quantum computing. 

In recent years, quantum computers have received significant attention in many fields of physics, chemistry, material science, computer science, etc \cite{Bassman,Jager,Chertkov,Jerbi, Konopik}. One of the most important applications of a quantum computer is electronic structure calculation of molecular systems which is the heart of computational chemistry. The electronic structure calculations for many-electron systems can play an important role in chemistry, solid-state physics, material science and pharmacology \cite{Langhoff,Kohanoff}. Some recent theoretical and experimental studies have shown the potential of molecular simulation using quantum computers \cite{OMalley,Colless,McCaskey,Leontica}. 

The fundamental goal of molecular electronic structure problems is to solve the ground state energy of many-body molecules. Knowing the molecular ground state properties allows researchers to obtain information about stable structures of the molecules, properties of the molecules and the mechanism and kinetics of the reactions \cite{Hoggan}. 

The electronic structure calculation is a computational challenge due to the presence of electron-electron correlations in molecules. Consequently, the problem of the ground state energy  can be solved by approximate approaches \cite{Hu}. In the weak interaction between electrons, the Hartree-Fock (HF) method can be used, which is a mean-field approximation \cite{Slater}. When the interaction between electrons becomes strong, one can use classical methods such as Configuration Interaction (CI) \cite{Hoggan,Szabo} and Coupled Cluster (CC) \cite{Hofmann} to obtain more accurate results. Furthermore, the Full Configuration Interaction (FCI) is a linear variational approach that can give a numerically exact solution for the ground state energy of the molecules without any truncation of the CI method. For this reason, FCI provides benchmark results against which other approximation approaches can be compared. In this approach, the wave function is expanded with all Slater determinants that can be generated based on a given one electron basis function. Since the number of Slater determinants depends factorially on the number of electrons, the FCI is only used for small to medium molecules \cite{Knowles1,Knowles2}.

Molecular simulation on classical computers would have a computational cost that grows exponentially with the size of the system \cite{Troyer}. Therefore, it seems that the use of quantum computers in molecular simulations is inevitable, especially for many-electron systems. 

In recent years, Variational Quantum Algorithms (VQAs) have been developed as a leading strategy to gain quantum advantage in Noisy Intermediate-Scale Quantum (NISQ) era computers, especially to simulate complicated quantum systems or solve large-scale linear algebra problems. VQA uses the classical optimizer and parameterized quantum circuits to run on a quantum computer. Unlike quantum algorithms developed for the fault-tolerant era, this approach has the additional advantage of reducing the depth of the quantum circuit, thereby reducing noise \cite{Cerezo}. In this regard, the Variational Quantum Eigensolver (VQE) algorithm was developed in 2014 to estimate the molecular ground state properties on NISQ era computer \cite{Peruzzo}.  Nowadays, VQE has attracted much attention in quantum computing due to their relatively low circuit depth and robustness against noise \cite{Preskill,Sharma} which has led to successful studies for the simulation of small molecules on available quantum hardwares and simulators by reducing the number of qubits \cite{Kandala,Grimsley,Nam,Hempel,Zhang}. However, more facilities, simulators and hardware are needed to extend this approach to larger systems \cite{Fedorov}.

VQE is a hybrid quantum-classical algorithm to estimate the minimum eigenvalue of a Hamiltonian by exploiting the Rayleigh-Ritz variational principle \cite{Peruzzo,Tilly}. In VQE, the ground state trial wave function of a molecule, called ansatz, is built up with adjustable parameters and a quantum circuit is designed to realize this ansatz. Then, the expectation value of the Hamiltonian is measured on a quantum computer. Finally, the parameters of the ansatz are optimized iteratively on a classical computer. VQE uses both quantum and classical computers, so it requires fewer quantum hardware facilities compared to pure quantum algorithms, such as the Quantum Phase Estimation (QPE) algorithm \cite{Dorner}. 

Here, we implement VQE to calculate the ground state energy of molecules such as Trihydrogen Cation (H$^+_3$), which is one of the most abundant ions in the universe and is the simplest triatomic molecule, Hydroxide ion (OH$^-$) which acts as a base, a ligand and a catalyst, Hydrofluoric Acid (HF) which uses to make most fluorine-containing compounds such as pharmaceutical medications and Trihydridoboron, also known as Borane (BH$_3$) which is a very strong Lewis acid. In these molecules, the number of qubits has an increasing trend. Moreover, we use computational chemistry approaches such as UHF and FCI to find the ground state energy. Consequently, the results of the ground state energy obtained from VQE are compared to FCI methods, as benchmark energy. It is important to note, we have implemented the Python-based Qiskit software on local clusters for our VQE simulation.

In the remainder of the paper, we first review the VQE framework for molecular electronic structure calculations including the derivation of the second-quantized Hamiltonian, fermion to qubit transformation, construction of ansatz and general steps of VQE. Then, the ground state energies of H$^+_3$,  OH$^-$, HF and BH$_3$ molecules are calculated based on VQE, UHF and FCI. Finally, the results are discussed in the conclusion section. 
  
\section{Variational Quantum Eigensolver Formalism}

\subsection{Second-quantized Hamiltonian}
We start with the definition of Hamiltonian of a molecule consisting of $M$ nuclei and $N$ electrons in atomic units ($\hbar=1$) as 

\begin{align}
H&=-\sum_i \frac{\nabla^2_{\Rr_i}}{2M_i} -\sum_i \frac{\nabla^2_{\rR_i}}{2}-\sum_{i,j}\frac{Z_i}{\vert \Rr_i-\rR_j\vert}\nonumber\\
&+\sum_{i,j>i}\frac{Z_iZ_j}{\vert \Rr_i-\Rr_j\vert}+\sum_{i,j>i}\frac{1}{\vert \rR_i-\rR_j\vert}
\end{align}
where $\Rr_i$, $M_i$ and $Z_i$ are the position, mass and charge of the nuclei, respectively, and $\rR_i$ denotes the position of the electrons. This form of the Hamiltonian is often said the first-quantized representation of molecular Hamiltonian. It generally imposes the fermionic nature of the electron through an antisymmetric wave function \cite{Hempel,McArdle}. Some methods were developed to tackle this Hamiltonian in electronic structure problems on quantum computer \cite{Kassal,Berry,Babbush}. Here, we analyze this problem in the second-quantized form of Hamiltonian. 

For the Hamiltonian representation in the second-quantized formulation, we first use the Born-Oppenheimer approximation which is common in Quantum Chemistry. In this approximation, based on the fact that the nuclei are much heavier than the electrons, we neglect the nuclei motion. So, the nuclei in a system are considered to be the fixed classical point charges, while the coordinates of the electrons are dynamic \cite{Born}. Moreover, to convert the Hamiltonian into a computational problem, we select basis functions $\phi_i$ as electronic wave function. Molecular orbitals (MOs) can be constructed as a linear combination of atomic orbitals (LCAO). The atomic orbital basis functions are derived from hydrogen-like atomic orbitals. They are numerically optimized to match desired physical properties of various systems. Basis functions with different forms are used in computational chemistry. The most simple basis functions for computational works are the Slater Type Orbital-$n$ Gaussians, STO-$n$G basis sets. These functions are expressed as the sum of Gaussian functions and the original Slater-type orbitals to increase the efficiency of the calculations \cite{Hempel,McArdle}. 

The electronic Hamiltonian in the second-quantized form can be written as \cite{Szabo}

\begin{align}
H=\sum_{pq} h_{pq} a^\dagger_p a_q+\frac{1}{2}\sum_{pqrs} h_{pqrs} a^\dagger_p a^\dagger_q a_r a_s
\end{align}
where $a^\dagger_p$ and $a_p$ are the fermionic creation and annihilation operators associated with $p$-th fermionic mode or spin-orbital. The electron is excited into single electron orbital $p$ by $a^\dagger_p$ and de-excited by $a_p$. Regarding the electron's spatial and spin coordinates $\Xx_i\equiv (\rR_i,\sS_i)$, the scalar coefficients in Eq.(2) can be calculated by

\begin{align}
h_{pq}=\int \D {\Xx} \phi^*_p({\Xx}) \Big(\frac{-\nabla^2_{\rR_i}}{2}-\sum_i \frac{Z_i}{\vert \Rr_i-\rR_i\vert}\Big) \phi^*_q({\Xx})
\end{align}
\begin{align}
h_{pqrs}=\int \D {\Xx}_1 \D {\Xx}_2\frac{\phi^*_p({\Xx}_1)\phi^*_q({\Xx}_2)\phi_r({\Xx}_1)\phi_s({\Xx}_2)}{\vert \rR_1-\rR_2\vert}
\end{align}
where are the one-electron and two-electron integrals, respectively.

In the first-quantized formulation, the number of electrons is conserved by the operators. In the second-quantized one, the operators must have an equal number of creation and annihilation operators. Moreover, in the second-quantized formulation, the anti-symmetry properties of the wave function are applied through the anti-commutation relations of the fermionic creation and annihilations operators as follows \cite{Szabo}

\begin{align}
\lbrace a_p,a^\dagger_q\rbrace=a_pa^\dagger_q+a^\dagger_qa_p=\delta_{pq}\\
\lbrace a_p,a_q\rbrace=\lbrace a^\dagger_p,a^\dagger_q\rbrace=0
\end{align}

The first-quantized formulation of Hamiltonian (in Hilbert space) directly stores the
wavefunction, without utilizing any previous information about the molecule. In the second-quantized case (in Fock space), our information about the molecule such as the occupation of electrons in energy levels (from the lowest to the highest) and the spatial form of the orbitals would be used. This knowledge leads to a reduction in the qubits needed to simulate molecules. Due to the small number of qubits available in current quantum computers, most quantum computing studies in chemistry to date have used the second-quantized formulation. The second-quantization form, also known as occupation number representation, must be used in the VQE algorithm in order to describe a fermionic system using qubits. The resulting fermionic problem must then be translated to qubits. \cite{McArdle}. 

\subsection{Fermion to Qubit Transformation}

To implement the second-quantized form of the Hamiltonian on a quantum computer, we need to transform the fermionic Fock space to the qubit's Hilbert space. This mapping makes the creation and annihilation operators of the fermions described by the unitary operators on the qubit \cite{McArdle}. The important methods for this mapping are: Jordan-Wigner (JW) \cite{Jordan}, Bravyi-Kitaev (BK) \cite{Bravyi} and parity \cite{Seely} transformations. 

In this section, we will focus on the parity transformation because of its simplicity. We consider a molecule with $N$ electrons and $M$ spin-orbitals. 

In the parity transformation, we use the $p^{\thh}$ qubit ($q_p$) to store the parity information of the first $p$ modes ($f_p$) as follows

\begin{align}
\vert f_{M-1},f_{M-2},...,f_0\rangle \longrightarrow \vert q_{M-1},q_{M-2},...,q_0\rangle
\end{align}
where $q_p=\Big[\sum_i^p f_i\Big] (\Mod 2)$.

The creation and annihilation operators can be written as 

\begin{align}
a_p&=\X_{M-1}\otimes...\otimes \X_{p+1}\nonumber\\
&\otimes \big(Q_p\otimes \vert 0\rangle \langle 0\vert_{p-1}-Q^{\dagger}_p \otimes \vert 1\rangle\langle 1\vert _{p-1}\big)
\end{align}
\begin{align}
a^{\dagger}_p&=\X_{M-1}\otimes...\otimes \X_{p+1}\nonumber\\
&\otimes \big(Q^{\dagger}_p\otimes \vert 0\rangle \langle 0\vert_{p-1}-Q_p \otimes \vert 1\rangle\langle 1\vert _{p-1}\big)
\end{align}
where $Q=\vert 0\rangle\langle 1\vert=\frac{1}{2}(\X+i\Y)$ in which X and Y are Pauli gates. The operators in Eqs.(8) and (9) analyze the parity of the $(p-1)^{\thh}$ modes, and adjust $p^{\thh}$ qubit considering $Q_p$ and $Q^{\dagger}_P$. Consequently, the string of Pauli gates updates all qubits that store the parity of $p^{\thh}$ qubit.
For example, the fermionic Fock state $a\vert 001\rangle + b \vert 010\rangle +c \vert 100 \rangle $ transforms to corresponding qubit state $ a \vert 111 \rangle + b \vert 110 \rangle +c \vert 100\rangle $. Moreover, $a_0 \longrightarrow X_2 X_1 Q_0$ where $n_i=a^{\dagger}_i a_i$ is the fermionic number operator. For more details, please see \cite{McArdle}.

\subsection{Trotter-Suzuki Approximation}
 
The qubit wave function can be evolved in time by transforming the time evolution operator of the system, $\EXP(-itH)$ on a series of gates. In this regard, the Trotter-Suzuki approximation \cite{Suzuki} is widely used. This approximation means that the Hamiltonian can be expressed as a sum of subsets that easily simulate the Hamiltonian, and then the whole evolution can be approximated as a sequence of this simpler evolution. Based on Trotter-Suzuki approximation, $\EXP(-itH)$ is divided into small subsets which can be conveniently mapped to circuits in a quantum computer. In the first-order Trotter-Suzuki fomula, we have \cite{Hu}

\begin{align}
\EXP(-itH)&\approx  \Big[\prod_{s=1}^n \EXP (-itH_s/N)\Big]^N 
\end{align}
where $H=\sum_{s=1}^n H_s$ and $H_s$ are local terms which act on a small subset of the particles in the system and $N$ is the Trotter number. The second-order Trotter-Suzuki fomula that requires more operations for better approximation can be written as

\begin{align}
\EXP(-itH) \approx  \Big[\prod_{s=1}^n \EXP (-itH_s/2N) \prod_{s=n}^1 \EXP (-itH_s/2N)\Big]^N
\end{align}

 In this work, the first-order Trotter-Suzuki approximation is used and $N=1$ to reduce qubit operations. For more details about Trotter-Suzuki Approximation, please see Appendix.

\subsection{Ansatz}

An important challenge of the VQE algorithm to calculate electronic structure on NISQ computers is to prepare an appropriate ansatz that accurately estimates the ground state energy of the molecules and could  be implemented by a low-depth circuit on quantum computers. In recent years, many ansatzes have been developed in VQE to solve the ground state energy problem of the molecular systems. For more details, please see \cite{Hu}. Here, we have used Unitary Coupled Cluster Single and Double excitations (UCCSD) ansatz. In the following, we review UCCSD ansatz.

The Unitary Coupled Cluster (UCC) ansatz is a chemically inspired ansatz was widely used in VQE \cite{Taube}. The origin of UCC comes from a computational chemistry approach called Coupled Cluster (CC). This ansatz constructs a parametrized trial wave function based on excitations above the initial state and can be defined as \cite{McArdle,Hu}

\begin{align}
\vert \psi({\bm \theta}) \rangle=\EXP \big(T({\bm \theta})-T^\dagger ({\bm \theta})\big) \vert \psi_0 \rangle
\end{align}
where $\vert \psi_0 \rangle$ is an initial state such as Hartree-Fock state and $U(\bm \theta)=\EXP \big(T({\bm \theta})-T^\dagger ({\bm \theta})\big)$  is unitary. Moreover, $T({\bm \theta})$ is the coupled cluster excitation operator which can be defined as

\begin{align}
T({\bm \theta})=\sum_s T_s(\bm \theta)
\end{align}
where $s$ denotes the excitation level 1,2,... However, in practice it is limited to single and double excitations. Moreover, $T_1$ and $T_2$ can be written as 
\begin{align}
T_1({\bm \theta})=\sum_{i,k} {\theta}_{ik} a^{\dagger}_k a_i
\end{align}
\begin{align}
T_2({\bm \theta})=\sum_{i,j,k,k'} {\theta}_{ijkk'} a^{\dagger}_k a^{\dagger}_{k'}a_i a_j
\end{align}
where $i$ and $j$ denote occupied spin orbitals, $k$ and $k'$ show unoccupied spin orbitals and $\theta$ is a different values of parameters.

To implement on a quantum computer, a UCCSD operator must be converted into single and two-qubit unitary gates using the first-order Trotter-Suzuki approximation to reduce the circuit depth which is an important limiting factor in NISQ era devices \cite{Romero}. The gate model of quantum computation is realistically limited to having gates working on just a few qubits at a time \cite{Grimsley}. Then, the mapping methods (as discussed in Section B) are required to transform the exponentiated creation and annihilation operators into Pauli operators.

\begin{figure*}
\centering
\includegraphics[scale=0.36]{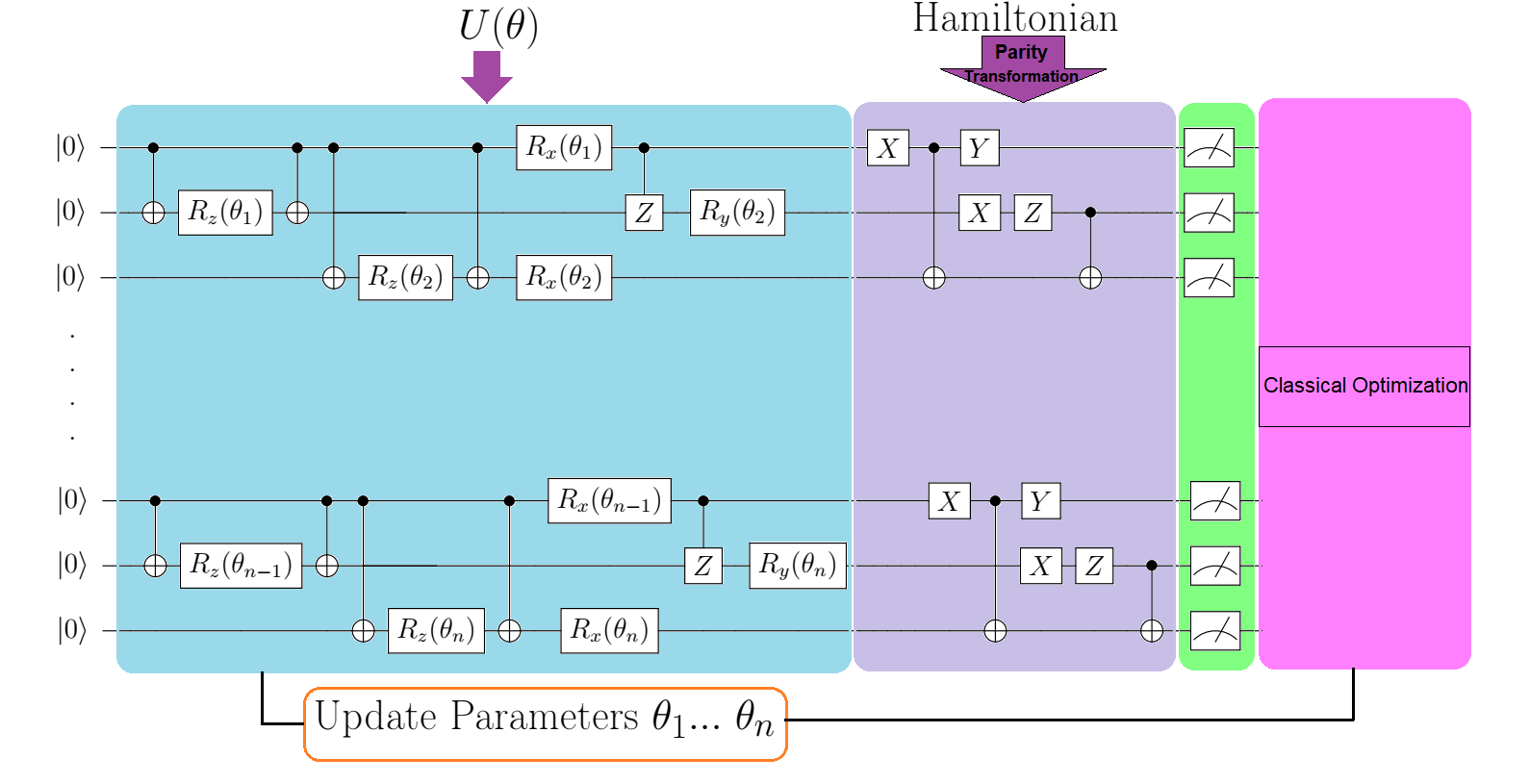}
\caption{A schematic of VQE to estimate the molecular ground state energy by adjusting variational parameters. The ansatz is prepared on a quantum computer as a quantum circuit consisting of gates such as $R_x(\bm \theta)$, $R_y(\bm \theta)$, $R_z(\bm \theta)$ and Z. The expectation value of the Hamiltonian is calculated by measuring every term of Hamiltonian on a quantum computer. Then, the energy and parameters are fed into the classical optimizer that updates the parameters for the next step of optimization. This procedure is performed repeatedly until the energy converges.}
\end{figure*}

\subsection{Variational Quantum Eigensolver}

VQE is the most promising algorithm in chemistry which is implemented on the NISQ devices. VQE estimates the ground state energy of the molecules using the Rayleigh-Ritz variational principle which is defined as \cite{Tilly}
\begin{align}
\langle \psi({\bm \theta})\vert H \vert  \psi({\bm \theta}) \rangle \geqslant E_g
\end{align}
where $E_g$ is the minimum eigenvalue of $H$ and $\psi(\bm \theta)$ is the ansatz for the ground state energy which is prepared by applying the parametrized gate $U(\bm \theta)$ on the initial state $\vert \psi_0\rangle$. Note that $E_g$ is the lowest model energy, not the true ground state energy. This principle states that one can find the ground state energy and wave function by optimizing the parameters $\bm \theta$  for a
particular geometric configuration which minimizes the expectation value of $H$.

The steps of the VQE are as follows \cite{Hu, Hempel}

1. The second-quantized Hamiltonian for the molecular systems should be obtained based on the computational basis set and solving one and two-electron integrals. Then, fermion to qubit transformation for the second-quantized Hamiltonian should be performed.

2. Based on a good classical approximation to the ground state of $H$ such as Hartree-Fock functions for the target system, an ansatz state $\psi(\bm \theta)$ would be prepared  by implementing unitary gates $U(\bm \theta)$ on a quantum circuit as $\psi(\bm \theta)=U(\bm \theta) \vert \psi_0\rangle$. The gates that operate on the qubits can be any parametrized gates such as single qubit rotations $R_x(\bm \theta)$, $R_y(\bm \theta)$ and $R_z(\bm \theta)$ or non-parametrized gates such as X, Y, Z, CNOT.

3. The expectation value of Hamiltonian $\langle H \rangle= \langle \psi({\bm \theta})\vert H \vert  \psi({\bm \theta}) \rangle$ which includes tensor products of Pauli operators should be measured for the prepared ansatz $\psi(\bm \theta)$. This step is often called Hamiltonian averaging. The state preparation and measurement steps are repeated many times to estimate the expectation value of $H$.  

4. The final step is optimization. In this step, the parameters $\bm \theta$ in ansatz $\psi(\bm \theta)$ are updated iteratively to minimize $\langle H \rangle$. We assume this procedure converges after $n$ iterations. Therefore, the variational ground state energy and wave function would be obtained. 

It is worth noting that fermion to qubit mapping, state preparation and measurement steps are performed on a quantum computer and the steps of second-quantized Hamiltonian and optimization would be performed on a classical computer. For this reason, VQE is a hybrid quantum-classical algorithm. Briefly, in VQE, the energy and parameters obtained from the quantum computer are input to a classical computer which outputs new parameters and are then fed back to the quantum circuit. This process is repeated until the energy converges \cite{McArdle}. A schematic of VQE is given in FIG.(1).

\begin{figure*}
\centering
\begin{subfigure}[]{
\centering
\includegraphics[scale=0.22]{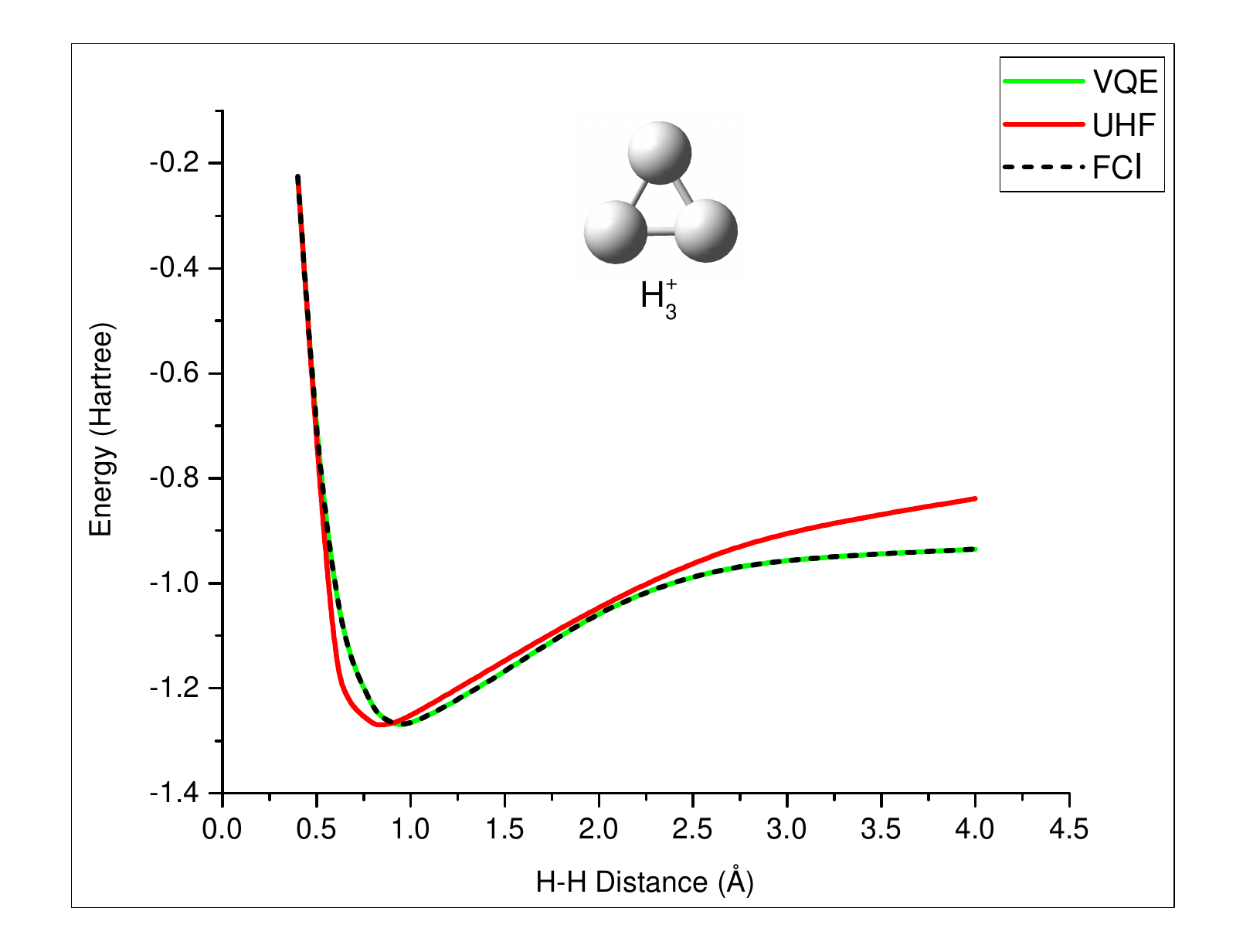}}
\end{subfigure}
\begin{subfigure}[]{
\centering
\includegraphics[scale=0.22]{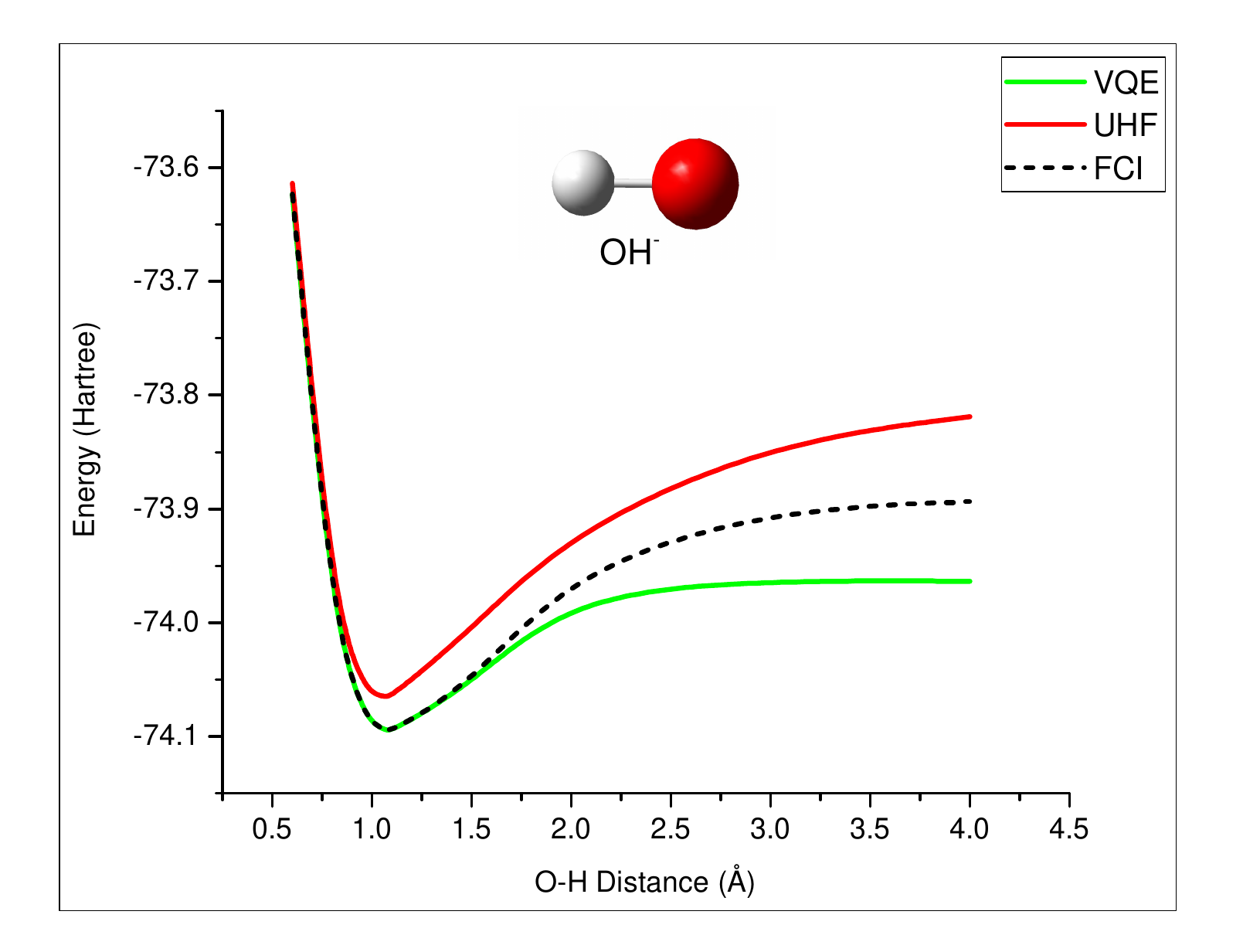}}
\end{subfigure} 
\begin{subfigure}[]{
\centering
\includegraphics[scale=0.22]{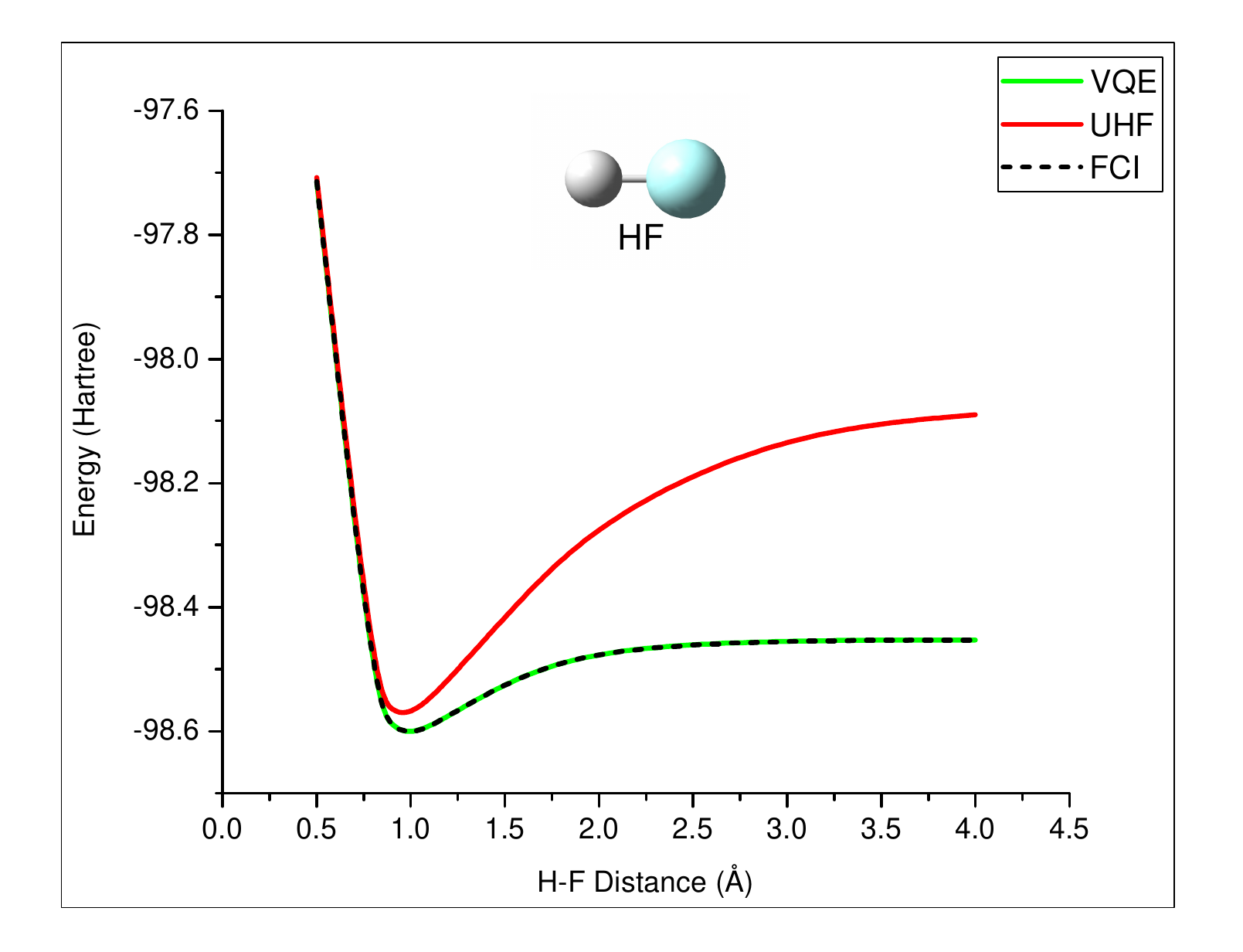}}
\end{subfigure}
\begin{subfigure}[]{
\centering
\includegraphics[scale=0.22]{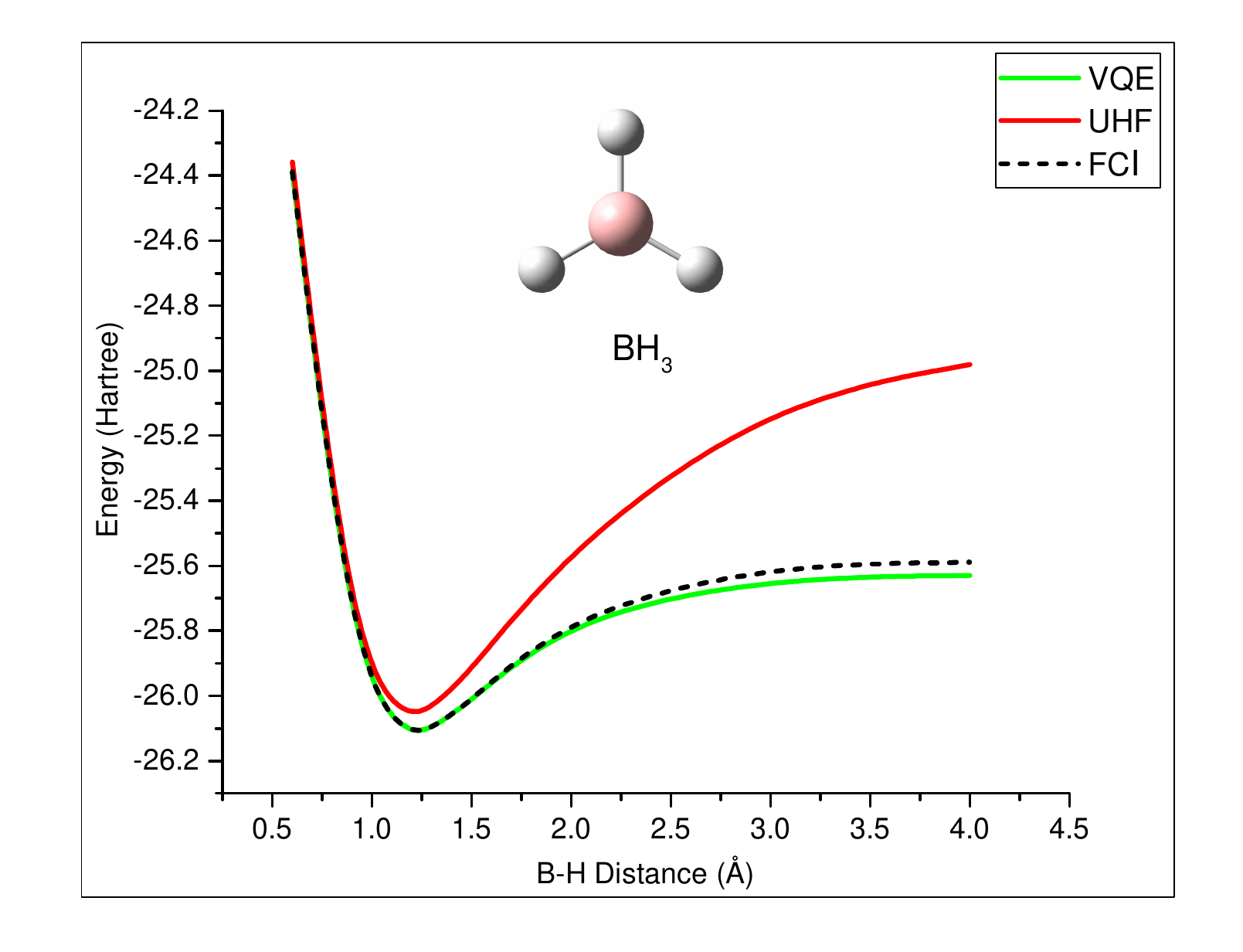}}
\end{subfigure}
\caption{Potential energy curves of a) H$^+_3$,  b) OH$^-$ c) HF and d) BH$_3$ molecules as a function of bond length (\AA) obtained from VQE, UHF and FCI approaches.}
\end{figure*}

\begin{figure*}
\centering
\begin{subfigure}[]{
\centering
\includegraphics[scale=0.154]{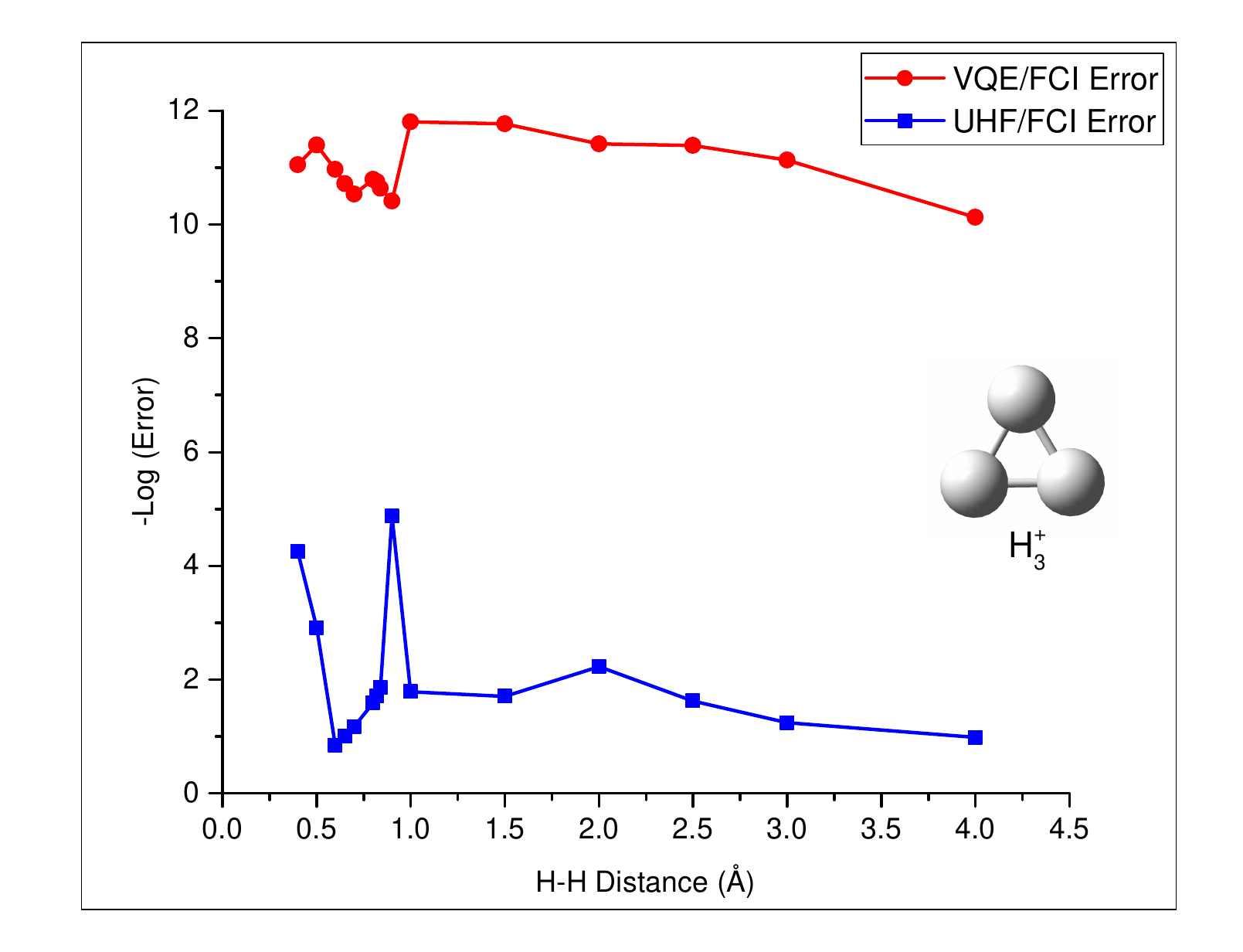}}
\end{subfigure}
\begin{subfigure}[]{
\centering
\includegraphics[scale=0.154]{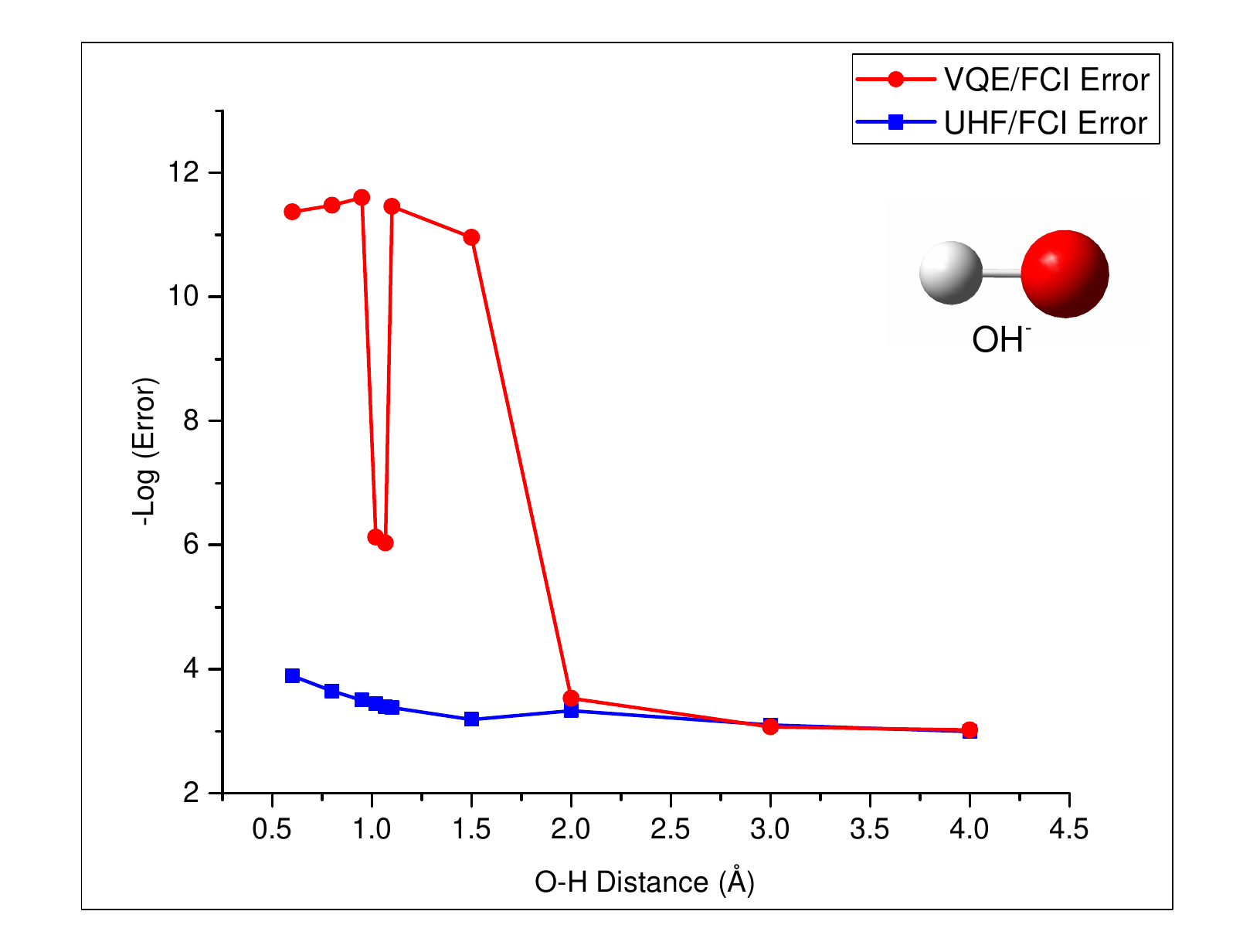}}
\end{subfigure} 
\begin{subfigure}[]{
\centering
\includegraphics[scale=0.154]{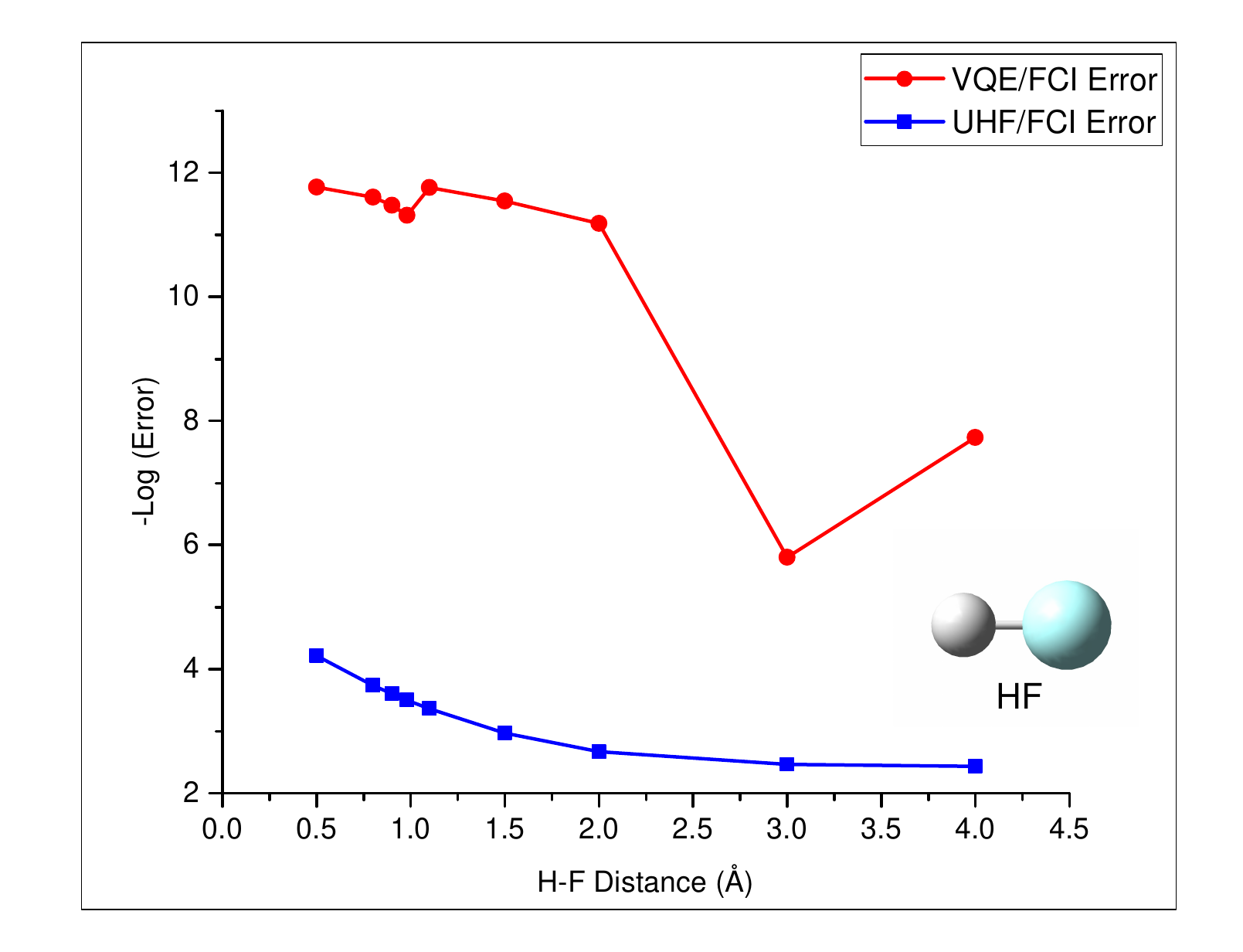}}
\end{subfigure}
\begin{subfigure}[]{
\centering
\includegraphics[scale=0.154]{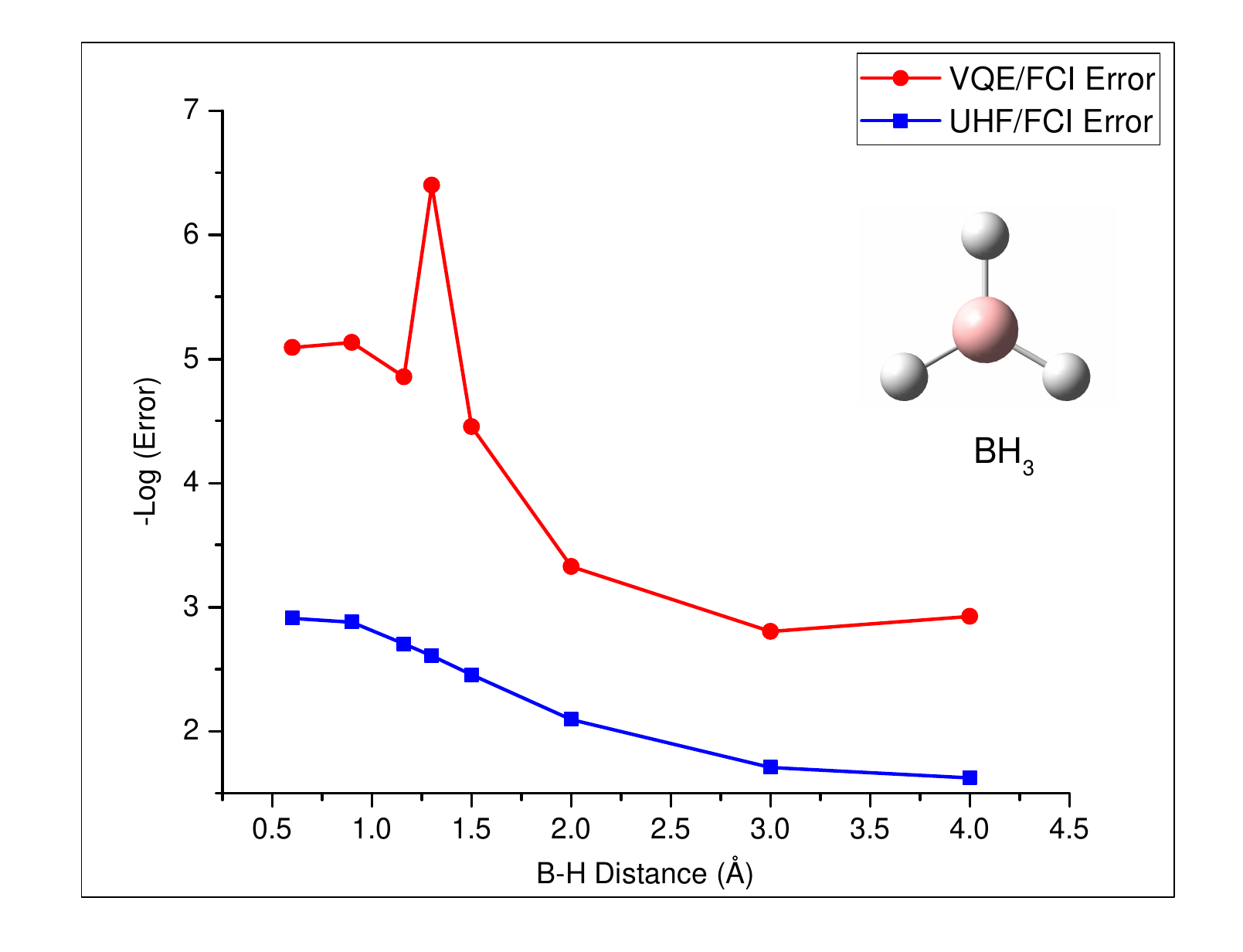}}
\end{subfigure}
\caption{Error in the ground state energy calculation of a) H$^+_3$, b) OH$^-$ c) HF and d) BH$_3$ molecules as a function of bond length (\AA) obtained from VQE and UHF in comparison to FCI energies as benchmark energy. As is clear, the energies obtained from the VQE algorithm are in better agreement with the FCI results than the UHF.}
\end{figure*}

\section{Results and Discussion}

In this section, we report and discuss the results obtained from the implementation of the VQE algorithm to calculate ground state energy of H$^+_3$,  OH$^-$, HF and BH$_3$ molecules.

To obtain the second-quantized Hamiltonian, Eq.(2), as an example, for HF molecule we start by considering the 1s orbital for hydrogen atom and 1s, 2s, 2p$_x$, 2p$_y$ and 2p$_z$ orbitals for fluorine atom. This leads to 12 molecular orbitals considering spin. Consequently, the calculation of the Hamiltonian of the HF molecule requires the consideration of 12 spin-orbitals. In this work, the molecular orbitals are calculated from the unrestricted Hartree-Fock (UHF) method in the STO-3G basis functions. Therefore, the Hamiltonian of HF molecule in the second-quantized form can be written as

\begin{align}
H=\sum_{p,q=1}^{12} h_{pq} a^\dagger_p a_q+\frac{1}{2}\sum_{p,q,r,s=1}^{12} h_{pqrs} a^\dagger_p a^\dagger_q a_r a_s
\end{align}
where $h_{pq}$ and $h_{pqrs}$ integrals are defined in Eqs.(3) and (4). The numerical solutions for these integrals for our studied molecules are performed by the Python-based Simulations of Chemistry Framework (PySCF) package \cite{Sun1,Sun2}. To transfrom fermion to qubit, we have used the parity transformation, as discussed in II.B section. 

In any transformation method such as parity, the number of qubits is equal to the number of spin-orbitals considered. Therefore, the number of qubits for H$^+_3$,  OH$^-$, HF and BH$_3$ molecules are 6, 12, 12 and 16, respectively. In other words, each qubit typically denotes the population of a spin orbital. However, since the molecular Hamiltonian has symmetry, the wave function can be stored in a smaller Hilbert space. We can reduce the number of qubits for the VQE algorithm considering $\mathbb{Z}_2$ symmetry. In this work, we chose a two-qubit reduction, which leads to a reduction in the complexity of the operation and especially useful for NISQ-era computers. Moreover, we have used UCCSD ansatz in our calculations which is a well-known chemically inspired ansatz in which a Hartree-Fock initial state is used as a starting point for the ansatz. The classical optimization process is performed using the Limited-memory BFGS Bound (L-BFGS-B) optimizer \cite{Romero}.

We have calculated the electronic ground state energy of H$^+_3$,  OH$^-$, HF and BH$_3$ molecules in various bond lengths using VQE, UHF and FCI methods. It is important to note that the bond length between H atoms in H$^+_3$ molecule is equal. The UHF and FCI energies are calculated by ORCA and Python, respectively.

The potential energy curves of the molecules as a function of bond length (\AA) are plotted in FIG.(2). As is clear, the molecular ground state energies obtained from the VQE algorithm lead to appropriate results. In FIG.(3), the ground state energies of studied molecules as a function of bond length (\AA) obtained from VQE and UHF are compared to the FCI results. By analyzing the errors based on logarithms, we conclude that there is a good agreement between molecular ground state energy obtained from VQE and FCI approach, especially in the equilibrium bond length of the molecules. Even in the dissociation limit of H$^+_3$, HF and BH$_3$ molecules, the agreement between VQE and FCI energies is considerable. Strictly speaking, the energies obtained from the VQE algorithm as a quantum algorithm are in better agreement with the FCI results than the UHF as classical computational energies. Furthermore, we have estimated the average accuracy of the energies obtained from VQE relative to FCI energies. The average accuracy of VQE results in our study and other works are given in TABLE I. Since we did not use a significant reduction of qubits, the average accuracy for the VQE results in our study is higher than the values reported by others.

\begin{table}[h] 
\caption{\small The order of magnitude of accuracy for VQE results in different works.} 
\centering
\setlength{\tabcolsep}{22pt} 
\begin{tabular}{c c c c c c c c c c  c c c c}
\hline \hline
{\bf Molecule} & {\bf Accuracy} & {\bf REF}\ \\ 
\hline \hline
H$_4$ & $10^{-4}$& \cite{Hu} \\
(H$_2$)$_2$ & $10^{-3}$& \cite{Zhang} \\
N$_2$ & $10^{-2}$& \cite{Zhang} \\
LiH & $10^{-3}$& \cite{Zhang} \\ 
H$_6$ & $10^{-3}$& \cite{Wu}\\ 
H$_{10}$ & $10^{-2}$& \cite{Wu}\\ 
H$_3$ & $10^{-9}$& \cite{Smart} \\ 
HeH$^{+}$& $10^{-3}$& \cite{Meitei} \\ 
H$^+_3$ & $10^{-10}$& This work \\ 
OH$^-$ & $10^{-7}$& This work \\
HF & $10^{-9}$& This work \\
BH$_3$ & $10^{-5}$ & This work \\
\hline \hline
\end{tabular}
\end{table}

\section{Conclusion}

Quantum computing is still a developing technology with limited computing capabilities due to the limited number of qubits available. One of the important applications of quantum computers is in theoretical chemistry in which the problem of molecular electronic structure and the estimation of the ground state energy play an important role. 

In recent years, VQE as a hybrid quantum-classical algorithm has been successfully developed to solve the electronic Schrödinger equation, especially for small molecules. In this work, we have used the VQE algorithm to calculate the molecular electronic structure and estimation of the ground state energy of some molecules such as H$^+_3$,  OH$^-$, HF and BH$_3$ in which the number of qubits is increasing and we have not used methods that greatly reduce the number of qubits. In this regard, after deriving the second-quantized Hamiltonian of the molecules using the STO-3G basis set, we have used the parity transformation for fermion to qubit mapping. Then, we constructed the ansatz based on the UCCSD. Consequently, the ground state enrgies of the molecules are obtained from VQE implementation on Python-based software package on local clusters. Moreover, we obtained the ground state energies based on the UHF method which is a common approach in computational chemistry. To investigate the accuracy of our results, we have obtained the FCI energies as benchmark energies and compared our results to them for each molecule. By checking the errors of VQE and UHF relative to the FCI results, we have concluded that there is a good agreement between molecular ground state energy obtained from VQE in comparison to the UHF, especially in the equilibrium bond length of the molecules. Moreover, we showed that the average accuracy of the VQE results in our work is higher than the previously reported values.

This work has been done with the aim of benchmarking the VQE algorithm to calculate the electronic ground state energy for a new set of molecules that can be good candidates for molecular simulation on a real quantum computer.

\begin{acknowledgments}
The authors would like to thank Ali Hayeri for his constructive comments and helpful discussion.
\end{acknowledgments}

\subparagraph{Authors Contribution}

H.N. proposed the idea and performed quantum algorithm computations. H.N. and E.M. contributed to the computational chemistry simulations. All authors contributed to the development and completion of the idea, analyzing the results and discussions. H.N. wrote the manuscript and H.D.Y and M.A. commented on it.

\subparagraph{Conflict of Interest}

The authors have no conflicts to disclose. 

\subparagraph{Data Availability}

The data that support the findings of this study are available on request from the corresponding author.

\appendix*
\section{Trotter-Suzuki Approximation}

 In both the Fermion and Pauli representations, the terms of a Hamiltonian $H$ that represents a molecular electronic structure problem do not commute. An important technique for simulating non-commuting $H$ on quantum computers is the Trotter-Suzuki approximation.

 It is generally challenging to directly implement $\exp(-itH)$ into quantum gates, however, $H$ is often the sum of a large number of individual terms $H_s$ such that it is easy to find circuits for the $\exp(-itH_s)$. As a simple example, consider $H=aX+bY+cZ$, so each of the individual terms can be simulated by $R_X,R_Y$ and $R_Z$ gates \cite{Low}.

Considering the fermionic $H$ in Eq.(2), we define

\begin{align}
a^{\dagger}=\frac{X-iY}{2}
\end{align}

\begin{align}
a=\frac{X+iY}{2}
\end{align}
where $X$ and $Y$ are Pauli gates. So, Eq.(2) can be written as 

\begin{align}
H&=\sum_{pq}h_{pq} \sum_{\sigma_p \sigma_q \in \lbrace X,\pm iY\rbrace} \sigma_p \sigma_q \nonumber\\
&+ \sum_{pqrs}h_{pqrs} \sum_{\sigma_p \sigma_q \sigma_r  \sigma_s \in \lbrace X,\pm iY\rbrace} \sigma_p \sigma_q \sigma_r \sigma_s
\end{align}
where $h_{pq},h_{pqrs} \in \mathbb{R}$ are the coefficients of the Pauli strings $\sigma$ \cite{Kang}.

To implement $H$ on quantum circuits, we split $H$ to individual terms ($H_s$). For this purpose, we use Trotter-Suzuki formula as

\begin{align}
\exp(-i\sum_s^n H_st) =\prod_s^n \exp(-iH_st)+\mathcal{O}(n^2t^2)
\end{align}
where if $t \ll 1$, error is minimal. Otherwise, we need to Trotterize as

\begin{align}
\exp(-i\sum_s^n H_st) =\Big(\prod_s^n \exp(-iH_st /N) \Big)^N+\mathcal{O}(n^2t^2/N)
\end{align}
where $N$ is Trotter step size. By regarding $N \in \mathcal{O}(n^2t^2)$, the error is minimal. Consequently, if we can implement $\exp(-i \sigma_p \sigma_ q t)$ and $\exp(-i \sigma_p \sigma_ q \sigma_ r \sigma_ s t) $ for $\sigma \in \lbrace X,Y,Z,I \rbrace$ on quantum circuits, we can map $H$ to circuits in a quantum computer.

\end{document}